\documentclass[aps,prd,amsmath,amssymb,longbibliography,
nofootinbib,twocolumn,showpacs,preprintnumbers]{revtex4-2}
\pdfoutput=1

\usepackage[toc,page]{appendix}
\usepackage{graphics}      
\usepackage{graphicx}
\graphicspath{{../}}
\usepackage{epsfig}
\usepackage{epstopdf}
\usepackage{float} 

\usepackage[table]{xcolor}    
\usepackage{url}           
\usepackage{bm}            
\usepackage{mathrsfs}
\usepackage{breakurl}

\usepackage[colorlinks=true,linkcolor=blue,citecolor=blue,urlcolor=blue]{hyperref}
\usepackage{color}
\usepackage{soul}
\soulregister\cite7
\usepackage{tikz}
\usetikzlibrary{calc}

\usepackage{pgfplots}
\pgfplotsset{width=5.7cm,compat=1.9}
\pgfplotsset{yticklabel style={text width=2.5em,align=right}}
\pgfplotsset{every axis/.append style={thick}}
\pgfplotsset{compat=1.18}

\newcommand\be{\begin{equation}}
\newcommand\ee{\end{equation}}
\newcommand\nono{\nonumber}
\newcommand\bse{\begin{subequations}}
\newcommand\ese{\end{subequations}}
\newcommand\bea{\begin{eqnarray}}
\newcommand\eea{\end{eqnarray}}
\newcommand{\ba}{\begin{align}}
\newcommand{\ea}{\end{align}}

\newcommand{\pd}{\partial}

\newcommand\ringring[1]{%
  {%
   \mathop{\kern0pt #1}\limits^{%
     \vbox to-1.85ex{
       \kern-2ex %
       \hbox to 0pt{\hss\normalfont\kern.1em \r{}\kern-.45em \r{}\hss}%
       \vss %
     }%
   }%
  }%
}
\newcommand{\q}{\quad}
\newcommand{\qq}{\qquad}
\newcommand{\Lie}{\mathcal{L}} %
\usepackage{xspace}
\newcommand{\nrpy}{{\texttt{NRPy}}\xspace}
\newcommand{\superb}{{\texttt{superB}}\xspace}

\allowdisplaybreaks
\begin{document}

\title{Binary black holes with non-orthogonal reference metrics}

\author{Nishita Jadoo}
\affiliation{Department of Physics, University of Idaho, Moscow, ID 83843, USA}

\author{Zachariah B.~Etienne}
\affiliation{Department of Physics, University of Idaho, Moscow, ID 83843, USA}
\affiliation{Department of Physics and Astronomy, West Virginia University, Morgantown, WV 26506, USA}
\affiliation{Center for Gravitational Waves and Cosmology, West Virginia University, Chestnut Ridge Research Building, Morgantown, WV 26505, USA}

\begin{abstract}
To our knowledge, numerical relativity (NR) simulations in the covariant moving-puncture formalism have only been performed with orthogonal reference metrics. Here, we demonstrate a new general non-orthogonal reference 3-metric implementation. We first validate this reference metric implementation by numerically evolving the exact Kasner initial data with a fisheye-style reference 3-metric, demonstrating numerical results that converge to the exact solution with increasing grid resolution. Next, we validate our approach by evolving binary black holes (BBHs) in the moving-puncture formulation, from inspiral through merger and ringdown. Even with a large deformation in the reference geometry, introduced to allow a large change  in grid refinement to enable the simulation to span the relevant length scales, from high resolutions near the BHs to low resolutions in the wave zone, no pathologies are observed. Smooth waveforms for up to $\ell=8$ are extracted on a grid without mesh refinement boundaries. We demonstrate the flexibility of tuning structured grids through the reference metric by employing spheroidal modifications of the spherical fisheye map to reduce the number of grid points in the $z$ direction when the orbit is in the $xy$-plane. These results support the use of smoothly derefined structured grids for BBH evolutions. 
\end{abstract}

\maketitle

\section{Introduction}

The ground-based LIGO, Virgo, and KAGRA gravitational wave (GW) detectors have reported hundreds of compact-binary merger detections, the large majority from BBH mergers, together with binary neutron star mergers~\cite{GWTC5}. Significant upgrades to existing ground-based detectors are planned~\cite{LIGOEvolvingDetectors,VirgoRuns}. Next-generation ground-based observatories are also planned~\cite{CosmicExplorer,EinsteinTelescope}, while the space-based LISA mission has entered its construction phase~\cite{ESALISAConstruction}. The increased sensitivity of future GW detectors will result in substantially higher detection rates and signal-to-noise ratios (SNRs). In particular, the high SNRs of individual events will impose increasingly stringent accuracy requirements on GW models and the NR simulations used to construct them~\cite{Purrer2020,Jan2024}. Such high-fidelity observations will also enable more precise tests of gravity in the nonlinear merger regime, for which controlling waveform systematics will be crucial~\cite{Maggio2023,Jan2024}.

Sampling the strong spacetime curvatures near comparable-mass BHs requires about two orders of magnitude higher resolution than GWs. To ensure reasonable computational cost, NR simulations of BBHs require numerical grids that efficiently span these orders of magnitude in length scale. Typically, this is accomplished through adaptive mesh refinement (AMR). However, in the most commonly used formulation of general relativity in NR (the moving puncture approach), a sharp gauge feature propagates outward and reflects off successive AMR refinement boundaries, producing substantial Hamiltonian constraint violations in its wake~\cite{Etienne2024}. These reflections inject noise into the simulations, resulting in inconsistent convergence with increasing resolution and obscuring high-order GW modes \cite{Baker2005, Pollney2011, Etienne2014}. On the other hand, simulating BBH mergers on a single, structured grid in general non-Cartesian coordinates can enable smooth changes in resolution from the strong to the weak field, which avoids reflections caused by discontinuous refinement changes and can lead to more accurate waveforms.  Fisheye grids have been successfully employed for this purpose in BBH and neutron star evolutions~\cite{Baker2002, Campanelli2006b, Campanelli2006, Campanelli2007, Faber2007, Liu2008, Zilhao2014}. However, earlier moving-puncture implementations of fisheye grids used non-covariant BSSN formulations, which required adjustments to outer boundary conditions and an approximate treatment of the transformed shift evolution equation~\cite{Faber2007}.

Years after the original fisheye moving-puncture BBH simulations, Brown \cite{Brown2009} formulated the BSSN equations in a covariant form that allows the use of any background 3-metric. The covariant BSSN formulation with curvilinear reference metrics has been successfully evolved using tensor rescaling to remove coordinate singularities~\cite{Montero2012,Ruchlin2018}. In those works the reference metric is orthogonal in the chosen coordinate basis and tensor rescaling factors are the diagonal entries of the reference metric and their derivatives. To our knowledge, this work provides the first demonstration of stable covariant BSSN BBH evolutions, without tensor rescaling, in which the reference 3-metric is non-orthogonal in the coordinate basis used for evolution. We show that despite a large deformation in the reference geometry to allow a large change in refinement close to the strong gravity region, the Hamiltonian and momentum constraints exhibit convergence consistent with the expected order, and no reflections from regions of derefinement are visible in the inspiral waveforms. Compared to earlier fisheye grid evolutions, using a fisheye reference metric has the advantage that all evolution, gauge, and boundary condition equations are written in covariant form, so no ad hoc modifications are needed~\cite{Brown2009}.

Building on the smoothly stretched orthogonal grids of Ref.~\cite{Ruchlin2018}, our work serves to numerically demonstrate, with a non-orthogonal reference metric, that grid refinement of structured grids can be changed smoothly through the reference metric leading to a consistent set of discrete equations that avoids spurious reflections. This can provide a pathway to using numerically generated structured grids to achieve refinement where needed, and can potentially be made dynamical through regridding or through modifications of the shift vector.

Previous work has explored BSSN formulations using cosmological background or reference solutions~\cite{Giblin2017,Hoshino2025}. Our work is a first step toward future numerical evolutions with time-dependent and curved reference metrics that may be useful for evolving BHs and GWs on cosmological backgrounds.

The outline of this paper is as follows. We begin in Sec.~\ref{sec:basic_eq} by reviewing the covariant BSSN evolution equations and give examples of non-orthogonal reference metrics. We provide details on the numerical implementation in Sec.~\ref{sec:numerical_impl}. In Sec.~\ref{sec:results} we present our numerical results, first validating the implementation against the exact Kasner solution and then applying it to BBH evolutions with fisheye and spheroidal fisheye reference metrics. We conclude in Sec.~\ref{sec:concl}.

\section{Basic equations}
\label{sec:basic_eq}

We briefly review the evolution equations used for vacuum spacetime in our open-source code \superb/\nrpy \cite{Ruchlin2018,Jadoo2025} and provide details on non-orthogonal reference 3-metrics employed. Throughout this paper, we adopt geometrized units ($G=c=1$). Latin indices ($i,j,k,\dots$) denote spatial components with repeated indices implying summation.

\subsection{Covariant BSSN evolution equations}

Brown~\cite{Brown2009} rewrote the traditional BSSN equations in covariant form by allowing the determinant of the conformal metric to vary and introducing a new tensor quantity: the difference between the conformal connection and the reference connection, $\bar \Gamma^k_{ij} - \hat \Gamma^k_{ij}$. All evolved variables are then tensors. We employ the Lagrangian choice with  $\pd_t \bar \gamma = 0$  described therein. Although no background metric, associated with the time-independent background connection, is assumed in that formulation, we define a background metric, $\hat \gamma_{ij}$, which we refer to as the reference metric. We evolve the difference between the conformal metric and the reference metric, $\varepsilon_{ij}$, rather than evolving the conformal metric itself. Additionally, we assume a flat reference metric and therefore set the reference metric curvature terms appearing in \cite{Brown2009} to zero.

The spacetime line element in the ADM decomposition is:
\be
ds^2 = -\alpha^2 dt^2 + \gamma_{ij} (dx^i + \beta^i dt)(dx^j + \beta^j dt) \ ,
\ee
where $\alpha$ is the lapse function, $\beta^i$ the shift vector, and $\gamma_{ij}$ the spatial 3-metric. The conformal metric $\bar \gamma_{ij}$ is defined as
\be
\gamma_{ij} = e^{4\phi} \bar \gamma_{ij} \ ,
\ee
where $\phi$ is the conformal factor. As in \cite{Tichy2007, Marronetti2008}, we evolve $W = e^{-2\phi}$ for better numerical behavior near punctures. The variable $\varepsilon_{ij}$ is defined as 
\be
\bar \gamma_{ij} = \hat \gamma_{ij} + \varepsilon_{ij} \ .
\ee
Quantities computed using the conformal metric (e.g., the covariant derivative $\bar{D}_i$ and Christoffel symbols $\bar{\Gamma}^k_{ij}$) are denoted by a bar, while those associated with the reference metric (e.g., $\hat{D}_i$ and $\hat{\Gamma}^k_{ij}$) carry a hat. The difference between their connection coefficients gives the definitions:
\begin{eqnarray}
\Delta^k_{ij} &=& \bar \Gamma^k_{ij} - \hat \Gamma^k_{ij} \ ,\\[1mm]
\Delta^k &=& \bar \gamma^{ij} \Delta^k_{ij} \ .
\end{eqnarray}
The new BSSN variable, $\bar \Lambda^i$, is constrained to satisfy
\be
\mathcal{C}^i = \bar \Lambda^i - \Delta^i = 0 \ .
\ee
The trace-free part of the conformal extrinsic curvature is defined as
\be
\bar A_{ij} \equiv e^{ - 4\phi}\Bigl( K_{ij} - \frac{1}{3} \gamma_{ij} K\Bigr) \ ,
\ee
where $K_{ij}$ is the extrinsic curvature and $K = \gamma^{ij} K_{ij}$ its trace. The lapse-weighted normal evolution operator is defined as
\be
\pd_\perp \equiv \pd_t - \Lie_\beta \ ,
\ee
where $\Lie_\beta$ denotes the Lie derivative along the shift vector $\beta^i$.

With these definitions, the evolution equations for the BSSN tensorial quantities are:
\begin{eqnarray}
  \label{eq:metric_RHS_orig_final} %
  \partial_{\perp} \varepsilon_{ij} & = & \frac{2}{3} \bar{\gamma}_{ij} \Bigl(\alpha \bar{A}_{k}^{k} - \bar{D}_k \beta^{k}\Bigr) + 2 \hat{D}_{(i} \beta_{j)} - 2 \alpha \bar{A}_{ij} \ , \\
  \label{eq:Abar_RHS_orig_final} %
  \partial_{\perp} \bar{A}_{ij} & = & -\frac{2}{3} \bar{A}_{ij} \bar{D}_k \beta^{k} - 2 \alpha \bar{A}_{ik} {\bar{A}^{k}}_{j} + \alpha \bar{A}_{ij} K \nonumber \\ %
  & & + e^{-4\phi} \Bigl\{-2 \alpha \bar{D}_i \bar{D}_j \phi + 4 \alpha \bar{D}_i \phi\, \bar{D}_j \phi \nonumber \\
  & & \qquad + 4 \bar{D}_{(i}\alpha\, \bar{D}_{j)}\phi - \bar{D}_i\bar{D}_j\alpha + \alpha \bar{R}_{ij}\Bigr\}^{\mathrm{TF}} \ , \\ %
  \label{eq:W_RHS_orig_final} %
  \partial_{\perp} W & = & -\frac{1}{3} W \Bigl(\bar{D}_k \beta^{k} - \alpha K\Bigr) \ , \\
  \label{eq:K_RHS_orig_final} %
  \partial_{\perp} K & = & \frac{1}{3} \alpha K^{2} + \alpha \bar{A}_{ij} \bar{A}^{ij} \nonumber \\
  & & \qq - e^{-4\phi} \Bigl(\bar{D}_i \bar{D}^{i} \alpha + 2 \bar{D}^{i} \alpha\, \bar{D}_i\phi\Bigr) \ , \\
  \label{eq:Lambda_RHS_orig_final} %
  \partial_{\perp} \bar{\Lambda}^{i} & = & \bar{\gamma}^{jk}\, \hat{D}_j\hat{D}_k\beta^{i} + \frac{2}{3} \Delta^{i}\, \bar{D}_j \beta^{j} + \frac{1}{3} \bar{D}^{i}\bar{D}_j\beta^{j} \nonumber \\ %
  & & - 2 \bar{A}^{ij} \Bigl(\partial_j \alpha - 6 \alpha \, \partial_j \phi\Bigr) \nonumber \\
  & & \qq + 2 \alpha \, \bar{A}^{jk} \Delta^i_{jk} - \frac{4}{3} \alpha\, \bar{\gamma}^{ij}\, \partial_j K \ ,
\end{eqnarray}
where $\beta_j=\hat\gamma_{jk}\beta^k$ in the reference metric derivative term, and $\mathrm{TF}$ indicates that the enclosed expression is taken to be trace-free with respect to $\bar{\gamma}^{ij}$, the conformal Ricci tensor $\bar R_{ij}$ is computed with
\begin{eqnarray}
  \bar{R}_{i j} {} = {} && - \frac{1}{2} \bar{\gamma}^{k l} \hat{D}_{k} \hat{D}_{l} \bar{\gamma}_{i j} + \bar{\gamma}_{k(i} \hat{D}_{j)} \bar{\Lambda}^{k} + \Delta^{k} \Delta_{(i j) k} \nonumber \\
  && + \bar{\gamma}^{k l} \left (2 \Delta_{k(i}^{m} \Delta_{j) m l} + \Delta_{i k}^{m} \Delta_{m j l} \right ) \ .
\end{eqnarray}
The trace-free condition $\bar{A}_{ij} \bar{\gamma}^{ij} = 0$ is enforced dynamically via the $\bar A^k_k$ term in Eq.~(\ref{eq:metric_RHS_orig_final}).

With the standard 1+log slicing \cite{Bona1995} and Gamma-driver shift \cite{Alcubierre2003} conditions, the lapse $\alpha$, shift $\beta^i$, and an auxiliary variable $B^i$ are evolved with:
\begin{eqnarray}
\label{eq:gauge_eqns}
\partial_t \alpha &=& -2 \alpha K + \beta^j \partial_j \alpha \ ,\\[1mm]
\partial_t \beta^{i} &=& B^{i} + \beta^j \bar{D}_j \beta^i \ ,\\[1mm]
\pd_t B^{i} &=& \beta^j \bar{D}_j B^i
+ \frac{3}{4} \left(
\partial_t \bar{\Lambda}^{i}
- \beta^j \bar{D}_j \bar{\Lambda}^{i}
\right)
- \eta B^{i} \ ,
\label{eq:damping_eta_final}
\end{eqnarray}
where $\eta$ is a damping parameter \cite{Schnetter2010}. Note that $\pd_t \bar \Lambda^i$ on the right-hand side of Eq.~(\ref{eq:damping_eta_final}) is substituted using Eq.~(\ref{eq:Lambda_RHS_orig_final}). 

Additionally, with initial data satisfying $\bar \gamma = \hat \gamma$, the Lagrangian choice $\pd_t \bar \gamma = 0$ preserves this algebraic determinant constraint, where $\bar \gamma$ and $\hat \gamma$ are the determinants of the conformal and reference metrics, respectively. This constraint is enforced numerically at each Runge--Kutta stage by setting 
\be 
\bar{\gamma}_{ij} \to
\left(\frac{\hat{\gamma}}{\bar{\gamma}}\right)^{1/3} 
\bar{\gamma}_{ij} \ .
\label{eq:detgammabar_condition}
\ee 

Hamiltonian and momentum constraints are computed using \cite{Ruchlin2018, Baumgarte2013} and monitored during the evolution:
\bea
  \mathcal{H} &=& \frac{2}{3} K^{2} - \bar{A}_{i j} \bar{A}^{i j}  \nono \\
  & & \qq + e^{-4 \phi} \big(\bar{R} - 8 \bar{D}^{i} \phi \bar{D}_{i} \phi - 8 \bar{D}^{2} \phi \big) \ ,  \label{eq:Hconstraint}\\
  \mathcal{M}^{i} &=& e^{-4 \phi}  (\hat{D}_{j} \bar{A}^{i j} + 2 \bar{A}^{k (i} \Delta^{j)}_{j k}  \nono \\
  & & \qq + 6 \bar{A}^{i j} \partial_{j} \phi 
  - {2}/{3} \bar{\gamma}^{i j} \partial_{j} K) \ ,  \label{eq:Mconstraint}
\eea
where $\bar{R} = \bar{\gamma}^{i j} \bar{R}_{i j}$.

\subsection{Non-orthogonal reference 3-metrics}

The line element in the reference 3D space is 
\be
ds^2 = \hat \gamma_{ij} d\hat x^i d\hat x^j \ ,
\ee 
where the reference metric, $\hat \gamma_{ij}$, is assumed to be time-independent, flat and nonsingular in the coordinate basis. Other than these assumptions, we consider general reference metrics, without specializing to any particular choice in our implementation, and demonstrate this generality by employing different reference metrics, specified at the outset, without requiring any further changes to the implementation.

\subsubsection{Fisheye reference metrics}
\label{subsubsec:fisheye_maps}

One way to achieve smooth derefinement is to stretch and compress Cartesian grid lines. An example of such a deformation, known as fisheye coordinates, has long been used in NR for evolving BBHs \cite{Baker2002,Campanelli2006b,Campanelli2006,Campanelli2007} and matter \cite{Faber2007,Shibata2006,Liu2008,Zilhao2014}.

We denote the computational fisheye coordinates by $\hat x^i = \{\hat x,\hat y,\hat z\}$ and the physical Cartesian coordinates by $x_{\mathrm{Cart}}^i = \{x_{\mathrm{Cart}},y_{\mathrm{Cart}},z_{\mathrm{Cart}}\}$. We call the latter physical because they are the Cartesian coordinates in which the flat reference metric takes its standard Euclidean form, whereas the fisheye coordinates $\hat x^i$ label points on the uniformly spaced computational grid. In a fisheye map, the radial grid stretching $dr/d\hat r$, where $r^2 = x_{\mathrm{Cart}}^2+y_{\mathrm{Cart}}^2+z_{\mathrm{Cart}}^2$ and $\hat r^2 = \hat x^2+ \hat y^2+ \hat z^2$, changes smoothly within transition regions and remains approximately constant between well-separated transitions. These transitions are implemented using hyperbolic tangent functions, since $\tanh(x)\to-1$ for $x\ll0$ and $\tanh(x)\to1$ for $x\gg0$.

\def\XMIN{0}
\def\XMAX{6}
\def\YMIN{0}
\def\YMAX{3.5}
\def\PLOTSAMPLES{180}
\def\CURVEWIDTH{semithick}
\colorlet{GuideColor}{gray}
\def\GuideStyle{dotted}
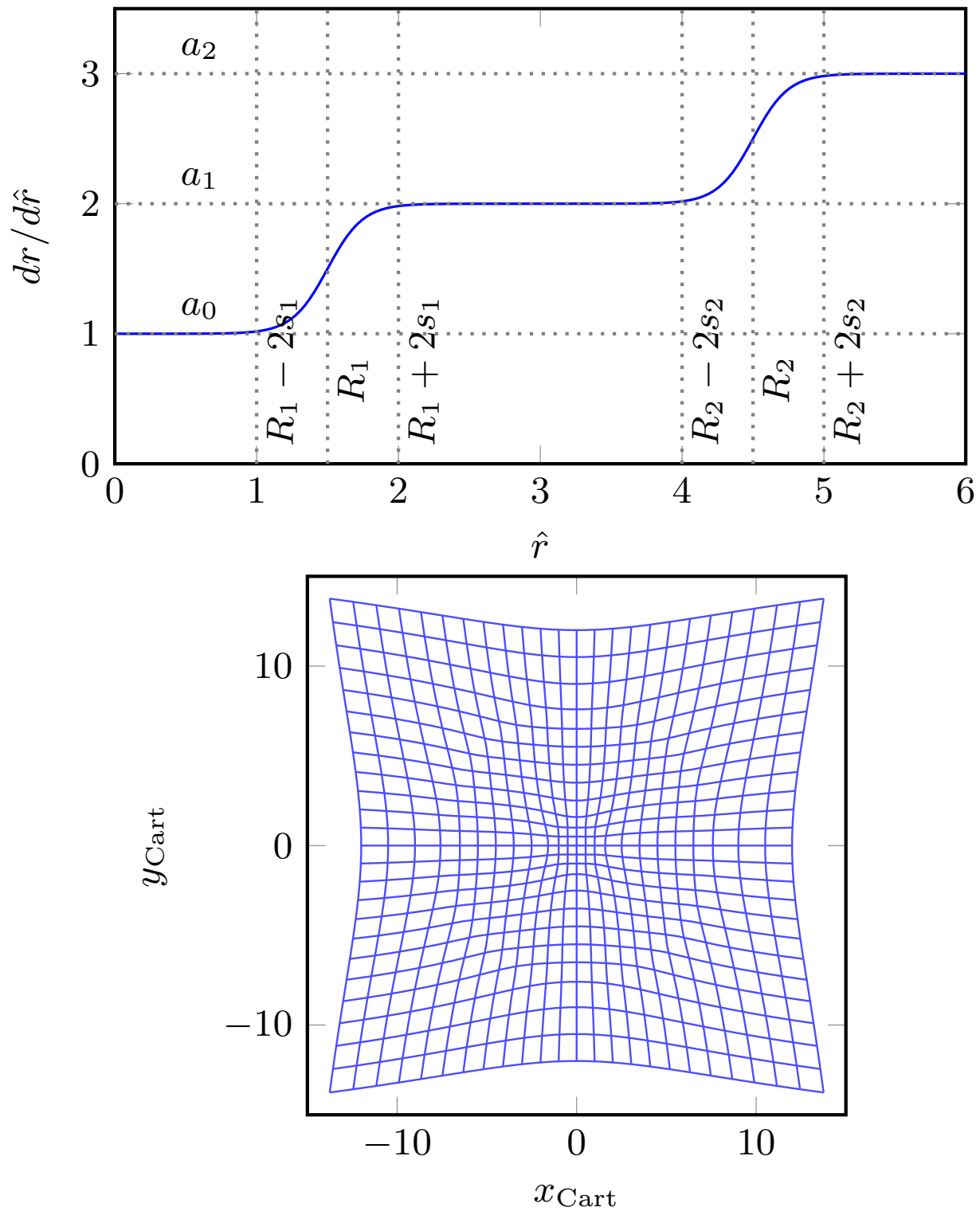
\begin{figure}
\centering
\pgfmathsetmacro{\aZero}{1.0}
\pgfmathsetmacro{\aOne}{2.0}
\pgfmathsetmacro{\aTwo}{3.0}
\pgfmathsetmacro{\ROne}{1.5}
\pgfmathsetmacro{\RTwo}{4.5}
\pgfmathsetmacro{\sOne}{0.25}
\pgfmathsetmacro{\sTwo}{0.25}
\pgfmathsetmacro{\cScale}{1.0}
\pgfmathsetmacro{\invTwoTanhOne}{
  (1+exp(-2*\ROne/\sOne))
  /(2*(1-exp(-2*\ROne/\sOne)))
}
\pgfmathsetmacro{\invTwoTanhTwo}{
  (1+exp(-2*\RTwo/\sTwo))
  /(2*(1-exp(-2*\RTwo/\sTwo)))
}
\pgfmathsetmacro{\coefOne}{
  (\aZero-\aOne)*\sOne*\invTwoTanhOne
}
\pgfmathsetmacro{\coefTwo}{
  (\aOne-\aTwo)*\sTwo*\invTwoTanhTwo
}
\pgfmathdeclarefunction{fetanh}{1}{%
  \pgfmathparse{
    sign(#1)*(1-exp(-2*abs(#1)))/(1+exp(-2*abs(#1)))
  }%
}
\pgfmathdeclarefunction{felncosh}{1}{%
  \pgfmathparse{
    abs(#1)+ln(1+exp(-2*abs(#1)))-ln(2)
  }%
}
\pgfmathdeclarefunction{ferbar}{1}{%
  \pgfmathparse{
    \cScale*(
      \aTwo*(#1)
      +\coefOne*(
        felncosh(((#1)+\ROne)/\sOne)
        -felncosh(((#1)-\ROne)/\sOne)
      )
      +\coefTwo*(
        felncosh(((#1)+\RTwo)/\sTwo)
        -felncosh(((#1)-\RTwo)/\sTwo)
      )
    )
  }%
}
\def\epsr{1e-8}
\pgfmathdeclarefunction{femappedcoord}{2}{%
  \pgfmathparse{
    (sqrt((#1)^2+(#2)^2)<\epsr)
    ? \cScale*\aZero*(#1)
    : ferbar(sqrt((#1)^2+(#2)^2))
      /max(sqrt((#1)^2+(#2)^2),\epsr)*(#1)
  }%
}
\begin{tikzpicture}
\begin{axis}[
  width=\columnwidth,
  height=0.62\columnwidth,
  xlabel={$\hat r$},
  ylabel={$dr/d\hat r$},
  ylabel style={at={(axis description cs:-0.06,.5)}},
  domain=\XMIN:\XMAX,
  xmin=\XMIN, xmax=\XMAX,
  ymin=\YMIN, ymax=\YMAX,
  samples=\PLOTSAMPLES,
]
\addplot[\CURVEWIDTH,blue]{
  \cScale*(
    \aTwo
    +(\aZero-\aOne)*\invTwoTanhOne*(
      fetanh((x+\ROne)/\sOne)
      -fetanh((x-\ROne)/\sOne)
    )
    +(\aOne-\aTwo)*\invTwoTanhTwo*(
      fetanh((x+\RTwo)/\sTwo)
      -fetanh((x-\RTwo)/\sTwo)
    )
  )
};
\addplot[\GuideStyle,GuideColor,no marks]
coordinates {(\XMIN,\aZero) (\XMAX,\aZero)}
node[pos=0.06,anchor=south west,text=black] {$a_0$};
\addplot[\GuideStyle,GuideColor,no marks]
coordinates {(\XMIN,\aOne) (\XMAX,\aOne)}
node[pos=0.06,anchor=south west,text=black] {$a_1$};
\addplot[\GuideStyle,GuideColor,no marks]
coordinates {(\XMIN,\aTwo) (\XMAX,\aTwo)}
node[pos=0.06,anchor=south west,text=black] {$a_2$};
\addplot[\GuideStyle,GuideColor,no marks]
coordinates {(\ROne,\YMIN) (\ROne,\YMAX)}
node[pos=0.10,anchor=west,xshift=6pt,rotate=90,text=black]
{$R_1$};
\addplot[\GuideStyle,GuideColor,no marks]
coordinates {(\RTwo,\YMIN) (\RTwo,\YMAX)}
node[pos=0.10,anchor=west,xshift=6pt,rotate=90,text=black]
{$R_2$};
\addplot[\GuideStyle,GuideColor,no marks]
coordinates {(\ROne-2*\sOne,\YMIN) (\ROne-2*\sOne,\YMAX)}
node[pos=0.10,anchor=west,xshift=6pt,yshift=-10pt,
rotate=90,text=black] {$R_1-2s_1$};
\addplot[\GuideStyle,GuideColor,no marks]
coordinates {(\ROne+2*\sOne,\YMIN) (\ROne+2*\sOne,\YMAX)}
node[pos=0.10,anchor=west,xshift=6pt,yshift=-10pt,
rotate=90,text=black] {$R_1+2s_1$};
\addplot[\GuideStyle,GuideColor,no marks]
coordinates {(\RTwo-2*\sTwo,\YMIN) (\RTwo-2*\sTwo,\YMAX)}
node[pos=0.10,anchor=west,xshift=6pt,yshift=-10pt,
rotate=90,text=black] {$R_2-2s_2$};
\addplot[\GuideStyle,GuideColor,no marks]
coordinates {(\RTwo+2*\sTwo,\YMIN) (\RTwo+2*\sTwo,\YMAX)}
node[pos=0.10,anchor=west,xshift=6pt,yshift=-10pt,
rotate=90,text=black] {$R_2+2s_2$};
\end{axis}
\end{tikzpicture}
\begin{tikzpicture}
\pgfmathsetmacro{\RawLim}{6.0}
\pgfmathsetmacro{\PlotLim}{15.0}
\pgfmathtruncatemacro{\GridDensity}{2}
\pgfmathsetmacro{\GridStep}{1/\GridDensity}
\pgfmathtruncatemacro{\GridN}{int(\RawLim*\GridDensity)}
\tikzset{
  mappedgrid/.style={line width=0.45pt,blue!70,no marks}
}
\begin{axis}[
  width=0.7\columnwidth,
  height=0.7\columnwidth,
  axis equal image,
  xlabel={$x_{\mathrm{Cart}}$},
  ylabel={$y_{\mathrm{Cart}}$},
  xmin=-\PlotLim, xmax=\PlotLim,
  ymin=-\PlotLim, ymax=\PlotLim,
  domain=-\RawLim:\RawLim,
  samples=28,
  clip=true,
]
\pgfplotsinvokeforeach{0,...,\GridN}{
  \pgfmathsetmacro{\k}{#1*\GridStep}
  \addplot[mappedgrid,variable=\t]
    ({femappedcoord(\k,\t)},{femappedcoord(\t,\k)});
  \addplot[mappedgrid,variable=\t]
    ({femappedcoord(\t,\k)},{femappedcoord(\k,\t)});
  \ifdim\k pt>0pt
    \addplot[mappedgrid,variable=\t]
      ({femappedcoord(-\k,\t)},{femappedcoord(\t,-\k)});
    \addplot[mappedgrid,variable=\t]
      ({femappedcoord(\t,-\k)},{femappedcoord(-\k,\t)});
  \fi
}
\end{axis}
\end{tikzpicture}
\caption{Top panel: Radial stretching $dr/d\hat r$ for a two-layer fisheye map with $a_0=1$, $a_1=2$, $a_2=3$, computational transition centers $R_1=1.5$, $R_2=4.5$, transition widths $s_1=s_2=0.25$, and normalization $c=1$. This derivative gives the local physical-to-computational spacing ratio along the Cartesian axes. Bottom panel: Uniform computational coordinate lines mapped into the physical $xy$-plane. The computational domain is $\hat x,\hat y\in[-6,6]$.}
\label{fig:drtildedr_fisheye_xy_lines}
\end{figure}

Several constructions of $dr/d\hat r$ using $\tanh$ functions exist in the literature (see, for example, Ref.~\cite{Zilhao2014}). Here, we use the profile given in Appendix~A of Ref.~\cite{Faber2007}, with a global normalization $c$:
\begin{equation}
\begin{aligned}
\frac{dr}{d\hat r}
={}&c\Bigg[
a_n+\sum_{j=1}^{n}
\frac{a_{j-1}-a_j}{2\tanh(R_j/s_j)}
\\
&\times\left\{
\tanh\!\left(\frac{\hat r+R_j}{s_j}\right)
-\tanh\!\left(\frac{\hat r-R_j}{s_j}\right)
\right\}
\Bigg] \ .
\end{aligned}
\label{eq:drtildedr}
\end{equation}
The two-layer example in Fig.~\ref{fig:drtildedr_fisheye_xy_lines} uses $c=1$. Integrating Eq.~\eqref{eq:drtildedr} with $r(0)=0$ gives
\begin{equation}
\begin{aligned}
r(\hat r)
={}&c\Bigg[
a_n\hat r+\sum_{j=1}^{n}
\frac{(a_{j-1}-a_j)s_j}{2\tanh(R_j/s_j)}
\\
&\times\ln\!\left(
\frac{\cosh((\hat r+R_j)/s_j)}
     {\cosh((\hat r-R_j)/s_j)}
\right)
\Bigg] \ ,
\end{aligned}
\label{eq:forward_fisheye_radial_map}
\end{equation}
and the physical Cartesian positions of the computational grid points are
\begin{equation}
x_{\mathrm{Cart}}^i=\left(\frac{r(\hat r)}{\hat r}\right)\hat x^i \ .
\label{eq:forward_fisheye_cartesian_map}
\end{equation}
The bottom panel of Fig.~\ref{fig:drtildedr_fisheye_xy_lines} shows the resulting computational coordinate lines in the physical $xy$-plane.

Without global normalization, the map generally changes the physical extent of the computational domain. Moreover, the transition centers and widths $R_j$ and $s_j$ are measured in the computational radius $\hat r$, rather than in physical radius. The global normalization used to enforce the desired physical grid extent, together with the conversion from physical input transition centers and widths $R_{j,\rm phys}$ and $w_{j,\rm phys}$ to the internal parameters $R_j$ and $s_j$, is described in Appendix~\ref{appendix:fisheye_scaling}.

We also employ an axisymmetric generalization of the radial fisheye map, which we refer to as spheroidal fisheye, with $x_{\mathrm{Cart}}=[r_\perp(\hat r)/\hat r]\hat{x}$, $y_{\mathrm{Cart}}=[r_\perp(\hat r)/\hat r]\hat{y}$, and $z_{\mathrm{Cart}}=[r_z(\hat r)/\hat r]\hat{z}$, where $r_\perp$ and $r_z$ use the forward radial map with their respective transition parameters and normalizations. 

\subsubsection{Inverse fisheye reference metrics}
\label{subsubsec:inverse_fisheye_maps}

We consider a fisheye construction for which the transition locations are prescribed directly in physical space to allow more precise tuning of the grid. In this construction, which we denote as inverse fisheye, the computational coordinates $\hat x^i$ are specified explicitly in terms of the physical Cartesian coordinates $x_{\mathrm{Cart}}^i$, rather than vice versa. We first define the spherical inverse map by
\begin{equation}
\begin{aligned}
\frac{d\hat r}{dr}
={}&c\Bigg[\frac{1}{a_n}
+\sum_{j=1}^{n}
\frac{a_{j-1}^{-1}-a_j^{-1}}
{2\tanh(2R_{j,\rm phys}/w_{j,\rm phys})}
\\
&\times\left\{
\tanh\!\left[\frac{2(r+R_{j,\rm phys})}{w_{j,\rm phys}}\right]
-\tanh\!\left[\frac{2(r-R_{j,\rm phys})}{w_{j,\rm phys}}\right]
\right\}\Bigg] \ .
\end{aligned}
\label{eq:inverse_fisheye_derivative}
\end{equation}
As in the forward construction, $a_j$ specifies the spacing relative to the central region. The normalization $c>0$ is computed internally from the prescribed physical grid extent, as described in Appendix~\ref{appendix:fisheye_scaling}. Integrating Eq.~\eqref{eq:inverse_fisheye_derivative} with $\hat r(0)=0$ gives
\begin{equation}
\begin{aligned}
\hat r(r)
={}&c\Bigg[\frac{r}{a_n}
+\sum_{j=1}^{n}
\frac{(a_{j-1}^{-1}-a_j^{-1})w_{j,\rm phys}}
{4\tanh(2R_{j,\rm phys}/w_{j,\rm phys})}
\\
&\times\ln\!\left(
\frac{\cosh[2(r+R_{j,\rm phys})/w_{j,\rm phys}]}
{\cosh[2(r-R_{j,\rm phys})/w_{j,\rm phys}]}
\right)\Bigg] \ ,
\end{aligned}
\label{eq:inverse_fisheye_radial_map}
\end{equation}
and the computational coordinates are
\begin{equation}
\hat x^i=\left(\frac{\hat r}{r}\right)x_{\mathrm{Cart}}^i \ .
\label{eq:inverse_fisheye_cartesian_map}
\end{equation}
Physical Cartesian coordinates $x_{\mathrm{Cart}}^i$ on the uniform computational grid are obtained by numerical inversion. This construction is specified independently and is not, in general, the exact inverse of the preceding forward map with unchanged parameter values.

We generalize this construction to spheroidal transition surfaces by introducing the layer-dependent physical ellipsoidal radii,
\begin{equation}
r_{{\rm ell},j}^{\,2}
=x_{\mathrm{Cart}}^2+y_{\mathrm{Cart}}^2+
\left(\frac{R_{j,\rm phys}}{R_{z,j,\rm phys}}\right)^2z_{\mathrm{Cart}}^2 \ ,
\label{eq:inverse_spheroidal_radius}
\end{equation}
where $R_{z,j,\rm phys}$ is the physical polar transition center. Dividing Eq.~\eqref{eq:inverse_fisheye_radial_map} by $r$, we replace $r$ by $r_{{\rm ell},j}$ in each term of the sum. We retain the Cartesian transformation in Eq.~\eqref{eq:inverse_fisheye_cartesian_map}, but the ratio $\hat r/r$ now depends on direction as well as physical radius. We refer to the resulting map as inverse spheroidal fisheye. The transition surface $r_{{\rm ell},j}=R_{j,\rm phys}$ has equatorial semiaxis $R_{j,\rm phys}$ and polar semiaxis $R_{z,j,\rm phys}$.

\section{Numerical implementation}
\label{sec:numerical_impl}

\superb/\nrpy already supported orthogonal reference metrics, such as Cartesian-like, spherical-like, and cylindrical-like metrics. The BSSN evolution and gauge equations, algebraic constraint~\eqref{eq:detgammabar_condition}, Sommerfeld boundary conditions, Kreiss--Oliger dissipation terms, moving-puncture improvement equations described in Ref.~\cite{Etienne2024}, and Hamiltonian and momentum constraint evaluations are implemented in covariant form and apply to any reference metric. Tensor rescaling is used to remove coordinate singularities associated with singular curvilinear coordinates using the diagonal entries of the reference metric and their derivatives as rescaling factors. 

We now extend the implementation to more general reference metrics. Since the coordinates are nonsingular, all tensor rescaling factors are set to unity and their derivatives to zero. We introduce a module that supplies the coordinate transformation. For the usual fisheye map, $x_{\mathrm{Cart}}^m=x_{\mathrm{Cart}}^m(\hat{x}^i)$ is given explicitly. For the inverse fisheye map, $\hat{x}^i=\hat{x}^i(x_{\mathrm{Cart}}^m)$ is instead explicit. In either case, \nrpy constructs the reference metric,
\be
\hat{\gamma}_{ij}=\delta_{mn}\frac{\partial x_{\mathrm{Cart}}^m}{\partial\hat{x}^i}\frac{\partial x_{\mathrm{Cart}}^n}{\partial\hat{x}^j}\ .
\ee
For greater computational efficiency, the provider may optionally supply algebraically equivalent expressions for the reference metric and its derivatives that are optimized for code generation and runtime evaluation.

Before constructing the initial data, the generated optimized C/C++ routine evaluates at every grid point and stores as auxiliary evolution grid functions the six independent components of $\hat{\gamma}_{ij}$, the 18 independent components of $\hat{\gamma}_{ij,k}$, and the 36 independent components of $\hat{\gamma}_{ij,kl}$. These quantities constitute 60 grid functions. Because the reference metric is time independent, they are evaluated once and reused throughout the evolution.

The existing initial data procedure, including the ADM-to-BSSN conversion, is unchanged. The added reference metric support supplies the coordinate map, its first derivatives, and the precomputed $\hat{\gamma}_{ij}$ and $\hat{\gamma}_{ij,k}$. The coordinate map determines the Cartesian position at which the initial data are evaluated, while its derivatives provide the required basis transformations. The precomputed reference metric fields are read from the auxiliary evolution grid functions when initializing the BSSN variables.

At each stage of the Runge--Kutta time integration, the existing covariant Ricci and right-hand-side routines read $\hat{\gamma}_{ij}$, $\hat{\gamma}_{ij,k}$, and $\hat{\gamma}_{ij,kl}$ from the auxiliary evolution grid functions. The constraint equations are also unchanged. When constraint diagnostics are requested, the Ricci routine first computes the conformal Ricci tensor using the reference metric and its first two derivatives. The routine that evaluates the constraints then directly reads $\hat{\gamma}_{ij}$ and $\hat{\gamma}_{ij,k}$ and uses the previously computed conformal Ricci tensor. Thus, $\hat{\gamma}_{ij,kl}$ enters the constraint evaluation through the conformal Ricci tensor.

As in the existing \superb/\nrpy implementation, we plot the momentum constraint magnitude as $\left|\mathcal{M}\right|\equiv\sqrt{\gamma_{ij}\mathcal{M}^i\mathcal{M}^j}$. For the binary runs, the radiation boundary conditions use fourth-order finite differences. The Kasner run uses quadratic extrapolation at the outer boundary.

\subsection{Comparison to previous fisheye grid implementations} 

Previous works employing fisheye grids are based on the non-covariant form of the BSSN equations (the covariant formulation had not yet been developed), in which the evolved fields are not tensors and no reference connection or metric is defined. At initial time, the evolved variables are transformed from Cartesian to fisheye coordinates (see Appendix~A of Ref.~\cite{Faber2007} for further details). In addition, the shift evolution and the radiation boundary conditions equations have to be modified. An exact transformation of these equations would make them more complex and increase computational cost, and hence approximations are employed. 

In comparison, in our reference metric implementation, no changes are needed to the evolution or boundary equations, making it straightforward to switch to different reference metrics at the outset. The shift evolution uses covariant derivatives, and the Sommerfeld boundary condition uses the radial derivative obtained through the coordinate map. There is no additional computational cost, during the evolution, in employing more complex reference metrics since no additional terms are needed in the evolution, boundary and constraint equations. 

The required storage of precomputed reference metric fields, such as $\hat{\gamma}_{ij}$, $\hat{\gamma}_{ij,k}$, and $\hat{\gamma}_{ij,kl}$, increases memory usage but avoids their repeated evaluation during the evolution, boundary-condition applications, and diagnostic calculations. It also avoids making the expressions involved in the latter more complicated than if the reference metric and its derivatives were expanded in terms of coordinates within those kernels.

\section{Results}
\label{sec:results}

In this section, we present validation results of the general reference metric implementation. Specifically, we perform the ``Apples-with-Apples'' Kasner test \cite{Vulcanov2002, Alcubierre2004} and reproduce the last BBH orbit simulation of Campanelli \textit{et al.}\ \cite{Campanelli2006} with a fisheye reference metric. We then evolve BBHs from a larger $8\,M$ initial separation with both spherical and spheroidal fisheye reference metrics. For the binary simulations, dimensional quantities are expressed in terms of $M$, where $M$ is the sum of the two target individual ADM masses, each equal to $M/2$.   

\subsection{Kasner ``Apples-with-Apples'' test}

\begin{figure}
    \centering
    \includegraphics{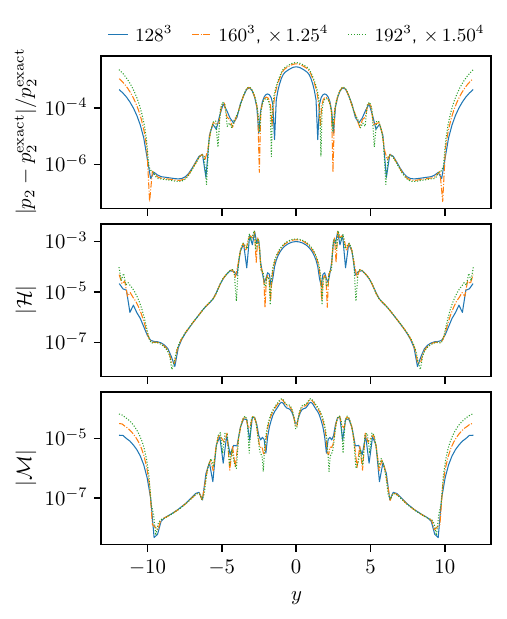}    
    \caption{Fractional error in the recovered Kasner exponent, $p_2$, the Hamiltonian constraint and the momentum constraint along the $y$ axis at $t\approx 4.95$. The $160^3$ and $192^3$ curves are rescaled for fourth-order convergence. The curves overlap, except at outer boundaries, as expected.}
    \label{fig:Kasner_ep2_H_and_M}
\end{figure}
The Kasner metric is an exact solution to the Einstein equations \cite{Kasner1921}:
\be
  ds^2 = -dt^2 + t^{2p_1} dx_{\mathrm{Cart}}^2 + t^{2p_2} dy_{\mathrm{Cart}}^2 + t^{2p_3} dz_{\mathrm{Cart}}^2 \ ,
  \label{eq:Kasner_metric}
\ee
where the Kasner exponents satisfy
\be
  p_1 + p_2 + p_3 = 1 \ , \qquad p_1^2 + p_2^2 + p_3^2 = 1 \ ,
  \label{eq:KasnerConditions}
\ee
which follow from the vacuum Einstein equations. In the Cartesian coordinate basis, the ADM quantities are:
\bea
  \alpha &=& 1 \ , \q
  \beta^i = 0 \ , \q \nonumber \\
  \gamma_{ij} &=& \operatorname{diag}\left(t^{2p_1}, t^{2p_2}, t^{2p_3}\right) \ ,
  \nonumber \\
  K_{ij} &=&
  -\operatorname{diag}\left(p_1 t^{2p_1-1}, p_2 t^{2p_2-1}, p_3 t^{2p_3-1}\right) \ .
  \label{eq:KasnerADM_quantities}
\eea
The Kasner metric and therefore the ADM quantities are spatially independent. The tensorial BSSN quantities when evolved with a Cartesian reference metric $\hat \gamma_{ij} = \delta_{ij}$ also have no spatial dependence. However, spatial dependence is introduced when using a fisheye reference metric.

Periodic boundary conditions are difficult to implement for a fisheye grid. We therefore use approximate extrapolation boundary conditions that do not preserve the constraints and stop the evolution before the outer boundary effects significantly contaminate the entire computational domain. 

The Kasner spacetime has no natural length scale, and units are omitted for all dimensional quantities. We evolve Kasner initial data with $(p_1,p_2,p_3)=\left(-1/3,2/3,2/3\right)$ at the initial time $t=1$ on a single-layer fisheye grid with an outer boundary at $12$ and fisheye parameters $a_0=1, a_1=2, R_{1,\rm phys}=3$ and $w_{1,\rm phys}=1$. We use fourth-order finite differences, fourth-order Runge--Kutta time integration, and a Courant factor of $0.5$, and evolve until $t=5$. We use a frozen lapse and zero shift for the gauge. We perform simulations at three grid resolutions, with $128^3$, $160^3$, and $192^3$ total grid points, using $68$, $85$, and $102$ iterations, respectively.

We recover $p_2$ from
\be
p_2^{\rm num}=-t\left(K^y{}_y\right)_{\rm Cart}\ .
\ee
The code reconstructs $\gamma_{ij}$ and $K_{ij}$ from the evolved fields, transforms them to the Cartesian basis, and forms $\left(K^i{}_j\right)_{\rm Cart}=\left(\gamma^{ik}\right)_{\rm Cart}\left(K_{kj}\right)_{\rm Cart}$. Figure~\ref{fig:Kasner_ep2_H_and_M} shows the fractional error in the recovered Kasner exponent, $p_2$, and the Hamiltonian and momentum constraints at $t\approx 4.95$ at grid points closest to the $y$-axis, with the higher resolution curves rescaled by the fourth-order convergence factor. The rescaled curves overlap closely, except near the outer boundary, where the constraints are not preserved. This overlap shows expected fourth-order convergence even at the fisheye transitions $y_{\mathrm{Cart}}=\pm 3$.

\subsection{Last orbit of BBHs}

\begin{figure}
\includegraphics{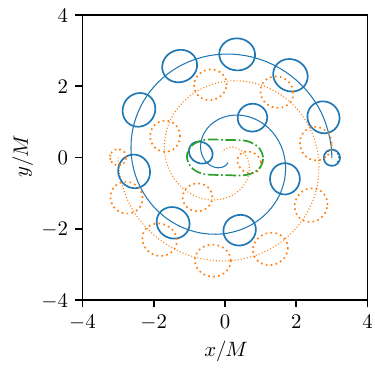}
\caption{Apparent horizons and center trajectories in the $xy$-plane for the $M/25$ run. Apparent horizons around each BH are plotted every $10\,M$ with solid blue and dotted orange lines. A common horizon is detected at $t \approx 113.5\,M$ and is shown with dash-dotted green lines at first detection only (cf.\ Fig.~1 of Ref.~\cite{Campanelli2006}).
\label{fig:last_orbit_horizons}}
\end{figure}

\begin{figure}
\includegraphics{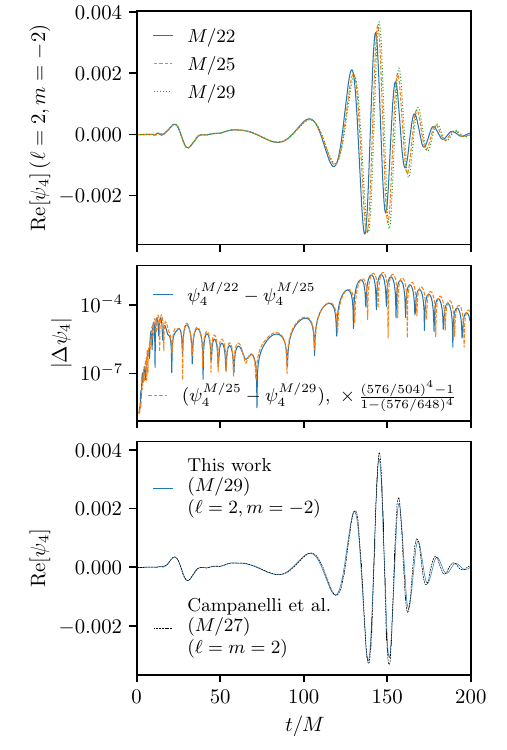} 
\caption{Top panel: Real part of the $(\ell=2,m=-2)$ mode of $\psi_4$ extracted at $r=20\,M$ for the three resolutions. Middle panel: Differences $\psi_4^{M/22}-\psi_4^{M/25}$ and $\psi_4^{M/25}-\psi_4^{M/29}$, with the latter rescaled for fourth-order convergence. Bottom panel: Overlay of $\mathrm{Re}[\psi_4]_{(\ell=2,m=-2)}$ from this work ($M/29$) and the $M/27$ curve digitized from Fig.~2 of Ref.~\cite{Campanelli2006}.
\label{fig:last_orbit_psi4}}
\end{figure}

\begin{figure}
 \includegraphics{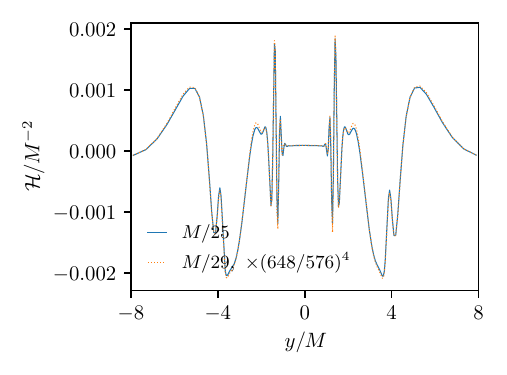}
\caption{The Hamiltonian constraint on the $y$-axis at $t\approx 70.01\,M$ for the $M/25$ and $M/29$ runs, with the latter rescaled for fourth-order convergence.  
\label{fig:last_orbit_H}}
\end{figure}

We reproduce the last orbit of the BBH simulation of Campanelli \textit{et al.}~\cite{Campanelli2006} using a single-layer fisheye grid similar to their run with the larger outer boundary. The goal is to evaluate the fisheye reference metric approach as an alternative to earlier fisheye grid implementations and the robustness of using a fisheye reference metric with aggressive fisheye scaling ($a_1=25$, $4\,M$ away from punctures) to evolve BBHs.

Other key differences with respect to Ref.~\cite{Campanelli2006} are that we use \texttt{TwoPunctures} instead of \texttt{BAM\_Elliptic} \cite{Brandt1997, Goodale2003} for initial data and our one-layer fisheye parameters closely approximate, but do not exactly match theirs. We place the BHs initially on the $x$-axis rather than the $y$-axis, which yields $\psi_{4,\ell,-m}=(-1)^\ell\psi_{4,\ell m}^{*}$, so the real parts of the $\ell=2,m=-2$ and $\ell=2,m=2$ modes agree. We therefore compare our $m=-2$ mode with the published $m=2$ mode. In both calculations, $M$ is the sum of the two target individual ADM masses, so no conversion of times or distances is required.

The initial data are equal-mass, nonspinning punctures constructed with \texttt{TwoPunctures} \cite{Ansorg2004}. We set the initial coordinate separation to $6\,M$, the tangential momentum to $p_t=0.13808$, and the radial momentum to $p_r=0$, with bare puncture masses $m_1=m_2=0.47656\,M$, \texttt{TwoPunctures} resolutions $(n_A,n_B,n_\phi)=(48,48,20)$ and initial lapse $\alpha=\psi_{\mathrm{BL}}^{-2}$, where $\psi_{\mathrm{BL}}=1+m_1/(2r_1)+m_2/(2r_2)$. We initialize the shift and Gamma-driver variable as $\beta^i=B^i=0$. The resulting total ADM mass reported by \texttt{TwoPunctures} is $0.984618\,M$, where $M/2$ is the target \texttt{TwoPunctures} ADM mass of each individual BH. The evolution uses the standard moving-puncture gauge, namely $1+\log$ slicing and the Gamma-driver shift, with damping parameter $\eta=2/M$. Hamiltonian constraint damping is disabled. We use the curvature-adjusted Kreiss--Oliger dissipation and moving-puncture techniques described in Ref.~\cite{Etienne2024}, with Kreiss--Oliger strengths $0.99$ for gauge variables and $0.3$ for non-gauge variables, and slow-start lapse (SSL) parameters $h=0.6$ and $\sigma=20\,M$. \footnote{We later found that SSL was not optimal for our runs because the conformal factor used to initialize the lapse omitted the regular correction $u$~\cite{Ansorg2004,Etienne2024}.}

The physical outer boundaries along the coordinate axes are located at $\pm 176.6\,M$. The fisheye reference metric is a single-transition map with physical parameters $a_0=1$, $a_1=25$, transition center $R_{\rm phys}=7\,M$ and transition width $w_{\rm phys}=8\,M$. We use three grids of sizes $504^3$, $576^3$, and $648^3$ points corresponding to central grid spacings of approximately
$M/22$, $M/25$, and $M/29$, respectively. Spatial derivatives are evaluated with fourth-order finite differences and time integration is performed with fourth-order Runge--Kutta. We use Courant factor $0.5$ and evolve to $t=200\,M$.

Apparent horizons are located during the evolution using the \texttt{BHaHAHA} library \cite{Etienne2026}. Figure~\ref{fig:last_orbit_horizons} shows the apparent horizons and horizon center trajectories for the two horizons at the start and the common horizon detected at $t\approx113.5\,M$ for the $M/25$ run. We search for a common horizon approximately every $0.25\,M$, and the reported time is the first successful detection. 

Figure~\ref{fig:last_orbit_psi4}, top panel, shows the real part of the $(\ell=2,m=-2)$ mode of $\psi_4$ extracted at $r=20\,M$ for the three resolutions $M/22$, $M/25$, and $M/29$. The waveforms are very similar across resolutions, with the dominant signal arising during the plunge and merger. The middle panel shows the differences $\psi_4^{M/22}-\psi_4^{M/25}$ and $\psi_4^{M/25}-\psi_4^{M/29}$, with the latter multiplied by the fourth-order factor, indicating approximate fourth-order convergence in the late inspiral and lower-order convergence during merger and ringdown. A small burst of spurious radiation is also present at early times. Our results are consistent with Ref.~\cite{Campanelli2006}. For a direct comparison with Ref.~\cite{Campanelli2006}, the bottom panel of Fig.~\ref{fig:last_orbit_psi4} overlays our $M/29$ waveform with the $M/27$ curve digitized from their Fig.~2. No amplitude rescaling, time shift, or sign change is applied. The two curves agree closely in phase and show similar amplitudes over the plotted interval.

Figure~\ref{fig:last_orbit_H} shows the Hamiltonian constraint along the $y$-axis at $t\approx 70.01\,M$, for the $M/25$ and $M/29$ runs. The higher resolution curve is rescaled by the fourth-order factor, and the close overlap of the two curves indicates excellent fourth-order convergence. The largest constraint violations occur near the punctures and fisheye transition region, while the outer part of the plotted interval remains comparatively small.

\subsection{BBHs from $8\,M$ separation}

\begin{figure*}
\includegraphics{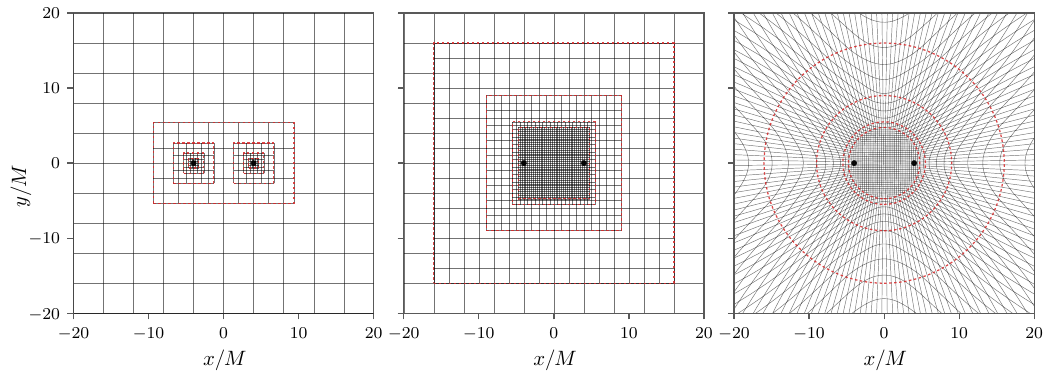}
\caption{Left panel: AMR grid for ET run. Middle panel: FMR grid with origin-centered boxes. Right panel: Eight-layer fisheye grid that achieves the same level of derefinement as the FMR grid smoothly. The punctures are plotted as solid circles. The red dashed lines plot the refinement boundaries for the AMR and FMR grids and the physical fisheye transition centers for the fisheye grid. Only every $16^{th}$ grid line is shown and the grid is plotted to $\pm20\,M$.}
\label{fig:grids_comparison}
\end{figure*}

After validating the general reference metric implementation, we next model a quasi-circular BBH from a larger separation to examine the quality of high $\ell$ waveforms from inspiral to ringdown for a fisheye single-grid. We demonstrate how the discrete changes in refinement of a possible fixed mesh refinement (FMR) configuration can be replaced with a similar continuous derefinement.

\begin{figure}
\centering
\includegraphics{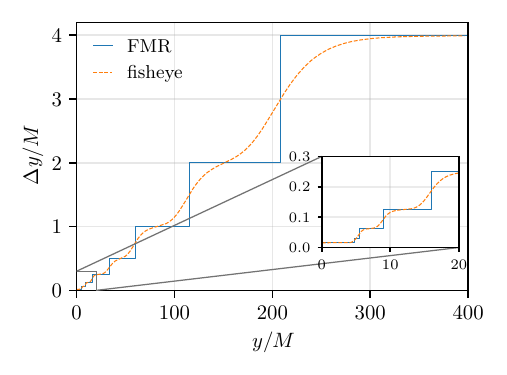}
\caption{ FMR grid spacing as a function of $y$ coordinate for the set of origin-centered  FMR boxes we consider in solid blue line and the eight-layer fisheye grid spacing that smoothly targets the  FMR spacing in orange dashed lines.}
\label{fig:8M_gridspacing}
\end{figure}

We consider fixed refinement since this paper focuses on fixed reference metrics. We set the outer boundary at $L = 400\,M$ and for a medium resolution grid consider the following plausible hierarchy of FMR boxes: centered on the origin with grid spacings  $\Delta x_{\mathrm{Cart},j}^{i}=\{M/64,M/32,M/16,M/8,M/4,M/2,M,2\,M,4\,M\}$ and half-widths
$\{4.75\,M,\allowbreak 5.5\,M,\allowbreak 9\,M,\allowbreak 16\,M,\allowbreak
34\,M,\allowbreak 60\,M,\allowbreak 115\,M,\allowbreak 208\,M,\allowbreak 400\,M\}$
(see middle panel of Fig.~\ref{fig:grids_comparison}.) To convert the FMR grid to a fisheye grid, we employ eight fisheye layers with the fisheye transition centers $R_{j,\rm phys}$ set to the half-widths of the FMR grid and fisheye scale factors to $a_j=\{1,2,4,8,16,32,64,128,256\}$, corresponding to the scale factors of the FMR grid spacings $\Delta x_{\mathrm{Cart},j}^{i}/\Delta x_{\mathrm{Cart},0}^{i}$ . The fisheye transition widths are set to $w_{j,\rm phys}= \{1\,M,\,1\,M,\,2\,M,\,4\,M,\,12\,M,\,20\,M,\,40\,M,\,80\,M\}$. To approximately match the FMR spacing between sufficiently separated transitions, the number of grid points in each computational direction $\hat x^i$ should be set to $N_{\hat x^i} = c (2L)/ \Delta x_{\mathrm{Cart},0}^{i}$, where $c$ is the global normalization factor described in Appendix~\ref{appendix:fisheye_scaling}. For our parameters this results in $N_{\hat x^i}\approx1388.9$, which we round to $1392$ for better grid partitioning. The fisheye grid is depicted in the right panel of Fig.~\ref{fig:grids_comparison}. Figure~\ref{fig:8M_gridspacing} shows FMR grid spacing along the $y$-axis and that of the eight-layer fisheye grid.

\begin{figure}
\centering
\includegraphics{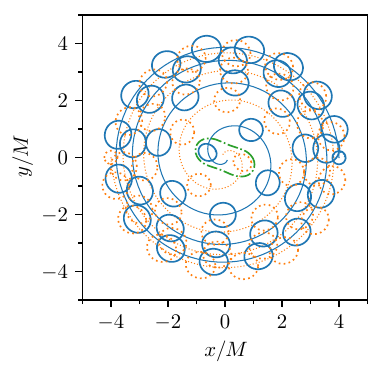}
\caption{Apparent horizons and center trajectories in the $xy$-plane for the high resolution run. Apparent horizons around each BH are plotted every $10\,M$ with solid blue and dotted orange lines. A common horizon is detected at $t \approx 378.4\,M$ and is shown with dash-dotted green lines at first detection only.
\label{fig:8Msep_horizons}}
\end{figure}

\begin{figure*}
\includegraphics{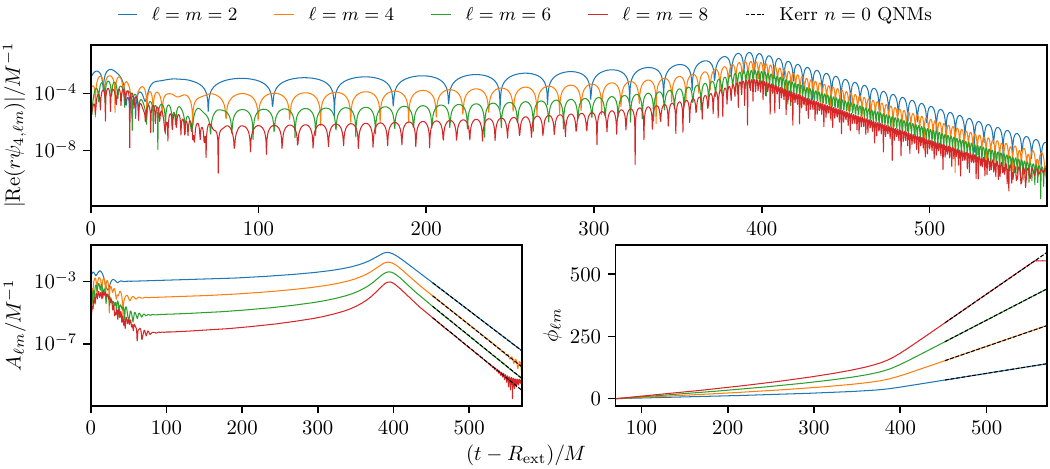}
\caption{Waveform modes $\ell=m=2,4,6,8$ extracted at radius $R_{\mathrm{ext}}=60\,M$ for the high resolution run. Top panel: Absolute value of the real part of $r\psi_{4,\ell m}$, showing the inspiral, merger, and ringdown frequencies. Bottom-left panel: Amplitude, $A_{\ell m}\equiv\left|r\psi_{4,\ell m}\right|$. Bottom-right panel: Phase computed from the real and imaginary components after the initial transient and shifted to zero at $t-R_{\mathrm{ext}}=70\,M$. Black dashed lines are plotted, starting $60\,M$  after the peak amplitude of the $\ell=m=2$ mode, for the predicted Kerr QNM amplitude and phase.}
\label{fig:8M_psi4}
\end{figure*}

As before, the initial data are equal-mass, nonspinning punctures constructed with \texttt{TwoPunctures}. We set the initial coordinate separation to $8\,M$ and choose the initial momenta using the 3.5PN quasicircular prescription implemented in \texttt{NRPyPN}~\cite{Habib2021}, with $p_t=0.1128452351$ and $p_r=-0.0022843438$. The target \texttt{TwoPunctures} ADM mass of each individual BH is $M/2$, and \texttt{TwoPunctures} obtains equal bare puncture masses $m_1=m_2=0.4823644225\,M$. We use \texttt{TwoPunctures} resolutions $(n_A,n_B,n_\phi)=(48,48,20)$ and initial lapse $\alpha=\psi_{\mathrm{BL}}^{-2}$. We initialize the shift and Gamma-driver variable as $\beta^i=B^i=0$. The resulting total ADM mass reported by \texttt{TwoPunctures} is $0.987917\,M$. The gauge, dissipation, and slow-start lapse parameters are unchanged from the last-orbit evolution.

The medium resolution grid consists of $1392$ points in each coordinate direction. For the medium resolution grid, we calculate the memory requirement as $8(1392)^3[4(24)+6+60]\approx 3.5\,\mathrm{TB}$, where 8 is the number of bytes per double-precision value and the terms in brackets account for four Runge--Kutta storage arrays for 24 evolved grid functions, six Ricci components, and 60 reference metric grid functions. This estimate excludes ghost zones and other allocations. This memory requirement indicates that the present single-grid approach is not a practical replacement for AMR, but serves to demonstrate smooth derefinement. For the low and high resolution runs, we scale the number of points in each direction by factors of $0.75$ and $1.25$, respectively, giving grids of $1044^3$ and $1740^3$ points. These correspond to central grid spacings of approximately $M/48$, $M/64$, and $M/80$ for the low, medium , and high resolution runs, respectively. Spatial derivatives are evaluated with eighth-order finite differences and time integration is performed with fourth-order Runge--Kutta. We use Courant factor $0.5$. 

The coordinate trajectories given by the horizon centers and the horizons computed by \texttt{BHaHAHA} are shown in Fig.~\ref{fig:8Msep_horizons} for the high resolution run. We search for a common horizon approximately every $0.25\,M$. The first common horizon detection occurs at $t=378.400\,M$ for the low, medium , and high resolution runs. From $t=0$ to $t=368.173\,M$, the maximum fractional changes in the two individual irreducible masses, $M_\textrm{irr}=\sqrt{A/(16\pi)}$, are $(2.47,3.32)\times10^{-5}$, $(2.62,3.33)\times10^{-5}$, and $(2.62,3.33)\times10^{-5}$ for the low, medium , and high resolution runs, respectively. We use the \texttt{BHaHAHA} stopping criteria $M\left\|\Theta\right\|_2<2\times10^{-5}$ and $M\left\|\Theta\right\|_\infty<10^{-2}$, where $\Theta$ is the outgoing-null expansion evaluated on the candidate horizon. The fractional changes in the individual irreducible masses do not decrease with increasing grid resolution, possibly because the \texttt{BHaHAHA} stopping tolerances are not sufficiently small.

We extract waveforms by decomposing the Weyl scalar $\psi_4$ into spherical harmonics of spin weight $-2$ up to $\ell=8$ at various radii. Figure~\ref{fig:8M_psi4} shows the real part, magnitude and phase of the $\ell=m=2,4,6,8$ modes of $\psi_4$ scaled by radius, extracted at $R_\mathrm{ext}=60\,M$. For an equal-mass, nonspinning binary, odd-$m$ modes vanish by $\pi$-rotation exchange symmetry, up to numerical symmetry-breaking errors. We create plots at a close radius of extraction since the radial grid spacing at this radius is small enough to resolve the $\ell=8$ modes and the smoothness of these modes can be demonstrated. The burst close to $t-R_\mathrm{ext}=0$ corresponds to initial junk radiation from relaxation of initial data. 

For the high resolution run, averaging the common horizon diagnostics over $433.577\,M\leq t\leq633.577\,M$ gives $M_{\mathrm{irr}}=0.884374\,M$ and $\chi_f=0.686332$. The peak-to-peak variation in $M_{\mathrm{irr}}$ is approximately $1.2\times10^{-5}\,M$. The corresponding Christodoulou mass, $M_f$, is $M_{\rm irr}/[(1+({1-\chi_f^2})^{1/2})/2]^{1/2}=0.951631\,M$. Evaluating the fundamental Kerr QNMs, $(s,\ell,m,n)=(-2,\ell,\ell,0)$, with the \texttt{qnm} package~\cite{Stein2019} at spin $\chi_f$, and converting from remnant-mass units to total-mass units using $M_f/M=0.951631$, gives the damping rates $\{\alpha_\ell\}_{\ell=2,4,6,8}=\{0.085425,\allowbreak 0.089162,\allowbreak 0.090545,\allowbreak 0.091127\}\,M^{-1}$ and frequencies $\{\omega_\ell\}_{\ell=2,4,6,8}=\{0.553417,\allowbreak 1.188253,\allowbreak 1.793647,\allowbreak 2.389485\}\,M^{-1}$. The corresponding Kerr QNM predictions are shown as black dashed lines over the ringdown portions of the bottom panels of Fig.~\ref{fig:8M_psi4}, with slopes equal to $-\alpha_\ell/\ln(10)$ for the logarithm amplitude plot and $\omega_\ell$ for the phase plot. The predicted damping rates and frequencies agree closely with the plotted ringdown behavior through $\ell=8$.

\begin{figure}            
\includegraphics{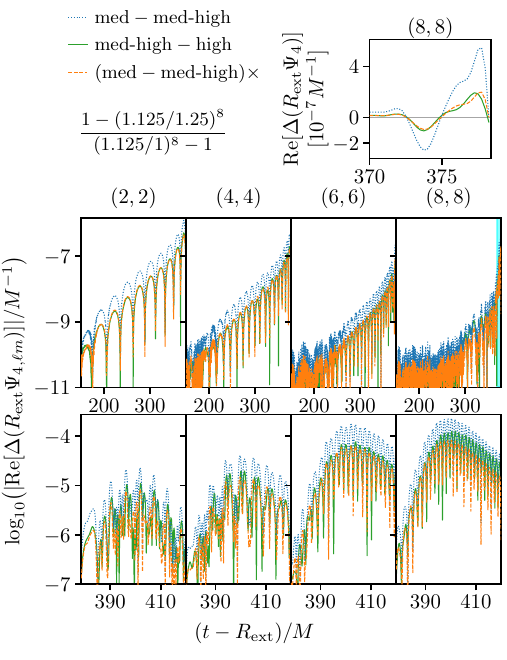}  
\caption{Convergence of the real parts of the $\ell=m=2,4,6,8$ waveform modes extracted at $R_{\mathrm{ext}}=60\,M$. The main panels show absolute differences between resolutions during inspiral (top) and merger to early ringdown (bottom). Orange dashed curves are rescaled for eighth-order convergence. The upper-right panel shows signed differences for $\ell=m=8$ over the cyan-shaded interval.}
\label{fig:8M_psi4_conv}
\end{figure}

We assess waveform convergence after the initial transient using differences in the real parts of the $\ell=m=2,4,6,8$ modes of $r\psi_4$ extracted at $R_{\mathrm{ext}}=60\,M$. Apparent convergence orders above eight involving the low resolution run suggested underresolution, prompting the addition of a medium high resolution run, scaling the number of points in each direction by a factor of $1.125$ relative to the medium resolution. Before computing differences between resolutions, we interpolate the real parts of the waveform modes onto the high resolution time grid using seven-point Lagrange interpolation. Figure~\ref{fig:8M_psi4_conv} shows the medium versus medium-high and medium-high versus high differences, with the former also rescaled for eighth-order convergence. The first row shows the inspiral till common horizon detection while the second row shows the merger to early ringdown. For the inspiral, the rescaled medium versus medium-high differences agree closely with the medium-high versus high differences, particularly when their magnitudes exceed approximately $10^{-9}\,M^{-1}$, consistent with eighth-order convergence. The signed differences for $\ell=m=8$, shown in the upper-right panel over the cyan-shaded interval, also agree closely in magnitude and oscillatory structure after rescaling. During merger and ringdown, the observed order is generally below eight, especially for the $\ell=m=8$ mode.

\begin{figure}
\includegraphics{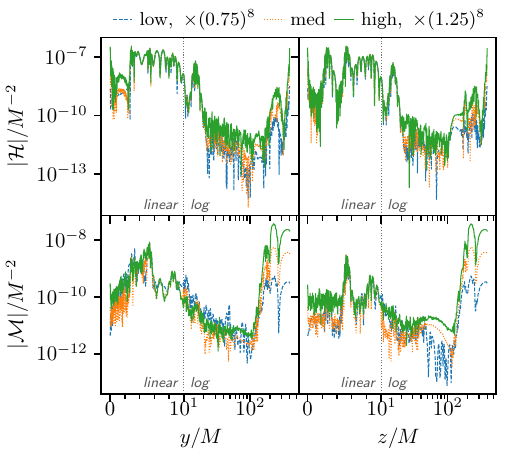}
\caption{Hamiltonian and momentum constraints on the $y$- and $z$-axes at $t\approx200\,M$ for the low, medium and high resolution runs, rescaled for eighth-order convergence. At $t \approx 200\,M$, the punctures have crossed the $y$-axis three times and are $\approx \pm 0.65\,M$ from the $y$-axis.}
\label{fig:8M_constraints}
\end{figure}

We plot the Hamiltonian and momentum constraints at three grid resolutions to verify the expected convergence order and assess the numerical accuracy of the simulation. Figure~\ref{fig:8M_constraints} shows the constraints along the $y$- and $z$-axes at $t\approx200\,M$, by which time the punctures have crossed the $y$-axis three times and lie approximately $0.65\,M$ on either side of it. The low and high resolution curves are rescaled relative to the medium resolution for eighth-order convergence. We observe approximately eighth-order convergence over large regions along the $y$- and $z$-axes. At $t\approx200\,M$, the high resolution constraints along the $y$- and $z$-axes satisfy $\left|\mathcal{H}\right|\lesssim10^{-7}\,M^{-2}$ and $\left|\mathcal{M}\right|\lesssim10^{-8}\,M^{-2}$. The highly deformed fisheye reference geometry compared to Cartesian is not observed to degrade convergence at the four inner fisheye transitions and does not appear to cause large constraint violations.   

\subsubsection{\texttt{Einstein Toolkit} AMR comparison}
\label{subsubsec:ET_comparison}

We reproduce the $8\,M$ initial separation BBH simulation using the open-source code
suite \texttt{Einstein Toolkit} (ET) \cite{ETK_2026} with AMR provided by the \texttt{Carpet} driver \cite{Carpet, Goodale2003, Schnetter2004}. The goal is to assess, for this specific BBH configuration, $\ell=m=8$ waveform quality improvements in the smoothly derefined fixed fisheye grid simulation over a discretely derefined AMR grid simulation with commonly used AMR parameters. However, the fixed fisheye grid is not a practical alternative to AMR as it involves much higher computational and memory costs. 

The ET grid has a coarse level and ten refinement levels with spacings $\Delta x_\ell=16\,M/2^\ell$, $\ell=0,\ldots,10$, and refined-level half-widths $\{680\,M,\allowbreak 340\,M,\allowbreak 170\,M,\allowbreak 85\,M,\allowbreak 42.5\,M,\allowbreak 21.25\,M,\allowbreak 5.375\,M,\allowbreak 2.6875\,M,\allowbreak 1.34375\,M,\allowbreak \tfrac{2}{3}\,M\}$. Refinement centers follow the punctures, initially at $x=\pm4\,M$, with overlapping boxes combined by Carpet. The outer boundary is approximately $1360\,M$ from the origin. The finest spacing is $M/64$. The grid spacing at $R_{\mathrm{ext}}=60\,M$ is $M$, and at $90\,M$, $120\,M$, and $150\,M$ it is $2\,M$. The inner grid differs from the fisheye grid (Fig.~\ref{fig:grids_comparison}).

We use \texttt{BaikalVacuum} with eighth-order finite differences, fourth-order Runge--Kutta integration, and \texttt{NewRad} radiative boundaries. Carpet uses fifth-order spatial and second-order temporal prolongation, with time-refinement factors $\{1,\allowbreak 1,\allowbreak 1,\allowbreak 1,\allowbreak 1,\allowbreak 2,\allowbreak 4,\allowbreak 8,\allowbreak 16,\allowbreak 32,\allowbreak 32\}$ and \texttt{Time::dtfac}$=1/64$. The \texttt{TwoPunctures} setup, initial lapse and shift, damping and dissipation strengths, and slow-start lapse parameters match the fisheye run. Hamiltonian constraint damping is also disabled. The shift equations use $\partial_j$ in place of $\bar D_j$ in Eqs.~\eqref{eq:gauge_eqns}--\eqref{eq:damping_eta_final}, and the tenth-difference Kreiss--Oliger operator has strengths multiplied by $W$. Waveforms are computed with eighth-order \texttt{WeylScal4} derivatives and decomposed through $\ell=8$ using \texttt{Multipole} with $(n_\theta,n_\phi)=(32,64)$ at the radii above.

\begin{figure}
\includegraphics{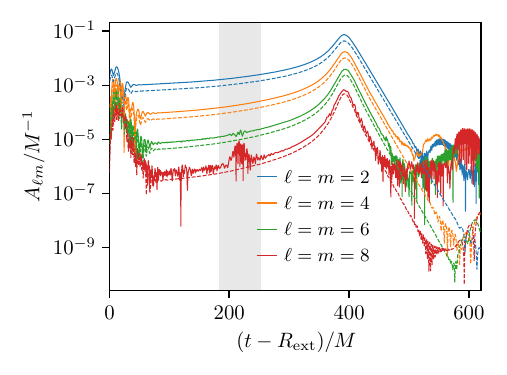}
\caption{Amplitudes of waveforms extracted at $R_{\mathrm{ext}}=60\,M$ for the medium resolution ET AMR (solid lines) and fisheye grid (dashed lines) runs. Dashed curves are shifted downward by 3 points for clarity. The grey shaded region indicates the estimated arrival interval for junk radiation reflected from the $85\,M$ AMR boundary.}
\label{fig:ET_psi4_ampl}
\end{figure}

A burst of small-amplitude, high-frequency oscillations is observed in the amplitudes of the ET waveforms for the $\ell=m=6$ and $\ell=m=8$ modes with peaks at retarded time $t-R_{\mathrm{ext}} \approx 216.2\,M$ at $R_{\mathrm{ext}} = 60\,M$, and is absent in the fisheye grid waveforms (see Fig.~\ref{fig:ET_psi4_ampl}). At the other extraction radii, $90\,M$, $120\,M$, and $150\,M$, the burst peaks occur at similar retarded times, approximately equal to $216.7\,M$, $217.5\,M$, and $218.1\,M$, respectively, suggesting a disturbance propagating outwards at the speed of light. 

One possible explanation for the burst is reflection of junk radiation from an AMR boundary, subject to the assumptions that $\psi_4$ mainly measures outgoing radiation and therefore ingoing junk radiation should not be significantly observed and that AMR reflections are much smaller when traveling from a coarser to a finer mesh and at finer meshes. Junk radiation originates close to punctures, peaking at $t\approx 13\,M$. The junk radiation reflects off an AMR boundary, travels inward, crosses to the opposite side and is detected at the extraction radius.  For reflection off an AMR boundary at $R_\mathrm{boundary}$, the estimated retarded arrival time ranges from $t-R_\mathrm{ext}=13\,M+2R_\mathrm{boundary}$ for an axial path to $t-R_\mathrm{ext}=13\,M+2\sqrt{2}R_\mathrm{boundary}$ for a face-diagonal path. For $R_\mathrm{boundary}=85\,M$, this gives an interval of approximately $183$--$253\,M$, shown as the gray shaded region in Fig.~\ref{fig:ET_psi4_ampl}. The expected arrival time of this reflection is consistent with the observed burst at all extraction radii. 

Figure~\ref{fig:ET_psi4_ampl} suggests approximate noise levels of order $10^{-6}\,M^{-1}$ for ET and $10^{-9}\,M^{-1}$ for the fisheye grid, with ringdown tails distinguishable for about $70$--$85\,M$ longer in the fisheye results. 

\subsubsection{Spheroidal fisheye modifications}
\label{subsubsec:spheroidal_fisheye}

\begin{figure}
\includegraphics{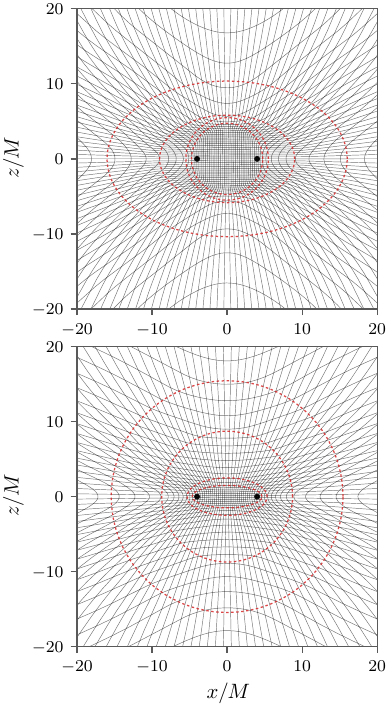}
\caption{Top panel: Spheroidal A grid modification. Bottom panel: Spheroidal B grid modification. Both grids are shown in the $xz$-plane. The punctures are plotted as solid circles, and the red dashed curves show the physical fisheye transition centers. Only every $16^{th}$ grid line is shown and the grids are plotted to $\pm20\,M$.}
\label{fig:spheroidal_fisheye_grid}
\end{figure}

We demonstrate the flexibility of tuning structured grids through the reference metric, employing two modifications of the eight-layer spherical fisheye grid, denoted spheroidal A and spheroidal B. Since the orbit lies in the $xy$-plane, we investigate whether fewer grid points can be used in the $z$ direction and compare the resulting waveforms with the spherical grid results. Spheroidal A increases the grid spacing in the $z$ direction everywhere outside the spherical region containing the orbit of the punctures whereas spheroidal B reduces this region of high resolution in the $z$ direction while keeping grid spacing outside of this region approximately unchanged.

Spheroidal A uses the spheroidal fisheye map described in Sec.~\ref{subsubsec:fisheye_maps}. Relative to the spherical fisheye grid, the $x$ and $y$ and first two $z$ fisheye transition parameters are unchanged. The $z$ transition radii and widths are multiplied by $0.648$ for layers 3--5 and by $0.72$ for layers 6--8. The grid in the $xy$-plane is therefore identical to that of the spherical fisheye configuration.

Spheroidal B uses the inverse spheroidal fisheye map described in Sec.~\ref{subsubsec:inverse_fisheye_maps}, with the same scaling factors $a_j=\{1,2,4,8,16,32,64,128,256\}$ as the spherical grid. The equatorial transition centres are approximately
$\{4.64094,\allowbreak 5.33048,\allowbreak 8.71745,\allowbreak 15.43789,\allowbreak 32.38184,\allowbreak 57.30117,\allowbreak 109.59777,\allowbreak 196.83502\}\,M$,
with width parameters
$\{4.43767,\allowbreak 5.53541,\allowbreak 1.90371,\allowbreak 3.80281,\allowbreak 11.45104,\allowbreak 19.19840,\allowbreak 38.89462,\allowbreak 74.89408\}\,M$.
The first two polar transition centres are $1.5\,M$ and $2.5\,M$ while the remaining polar centres equal their equatorial transition centers. Each layer uses the same width parameter in the polar and equatorial directions.

At the medium resolution grid scale, spheroidal A and B contain $1392\times1392\times1152$ and $1392\times1392\times1104$ points, respectively, reducing the point count in the $z$ direction by $17.24\,\%$ and $20.69\,\%$. Figure~\ref{fig:spheroidal_fisheye_grid} shows the $xz$-plane grids for both modifications.

\begin{figure}
\includegraphics{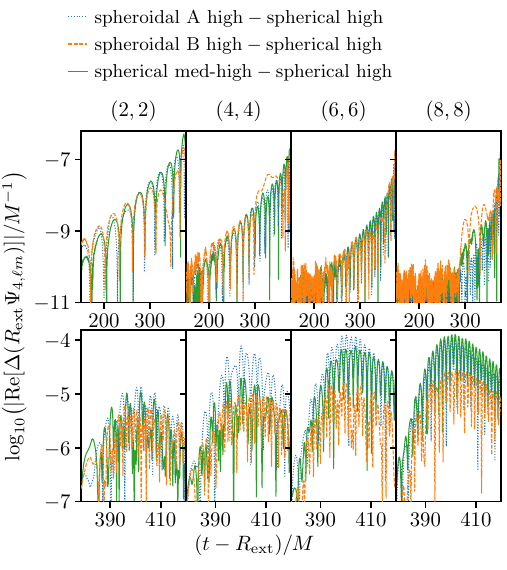}
\caption{Absolute differences in the real parts of the $\ell=m=2,4,6,8$ modes of $R_{\mathrm{ext}}\psi_4$ at $R_{\mathrm{ext}}=60\,M$ during inspiral (top) and merger to early ringdown (bottom). High resolution results for spheroidal A (blue dotted) and spheroidal B (orange dashed) are compared with the spherical high resolution run. Green solid curves show the spherical medium-high versus high differences.}
\label{fig:8Msep_psi4_spherical_vs_spheroidal}
\end{figure}

Figure~\ref{fig:8Msep_psi4_spherical_vs_spheroidal} compares the high resolution waveforms obtained with spheroidal A and B against the spherical high resolution run. We also show the spherical medium-high versus high differences. The medium-high spherical run uses $10\,\%$ fewer points in each coordinate direction than the high resolution spherical run. During inspiral, spheroidal A produces smaller differences from the spherical high resolution waveforms than spheroidal B, particularly for $\ell=m=4$ and $8$. This suggests that retaining fine resolution over a larger region around the binary, as in spheroidal A, helps resolve the inspiral. During merger and early ringdown, spheroidal B generally produces smaller waveform differences.

\section{Conclusions}
\label{sec:concl}

In this work, we have numerically demonstrated stable BBH evolutions of the covariant BSSN equations with non-orthogonal, nonsingular, flat, time-independent reference 3-metrics, without tensor rescaling. We first showed convergence for an exact solution of the Einstein equations, the Kasner solution in Cartesian coordinates, for which using a non-Cartesian reference metric adds spatial dependence for the evolved BSSN tensorial quantities. We then reproduced the last-orbit BBH simulation of Ref.~\cite{Campanelli2006} using a comparable strong fisheye derefinement implemented through a fisheye reference metric, obtaining similar results. 

We simulated BBHs from a larger $8\,M$ initial separation using several fisheye layers that achieve the same level of derefinement as FMR boxes centered on the origin. We obtain smooth waveforms up to $\ell=8$ on a grid without mesh refinement boundaries. Using spheroidal fisheye reference metrics, we demonstrated the flexibility of reference metrics in modifying structured grids.

The general reference metric implementation in the open-source NR code \nrpy will facilitate the exploration of structured grids suited to specific problems as well as their dynamical extension through modifications of the shift vector and regridding. Furthermore, building on our validated implementation of time-independent reference metrics, we plan to extend the covariant BSSN implementation to time-dependent reference metrics, to evolve BHs and GWs on time-dependent cosmological backgrounds.

\begin{acknowledgments}
N.J. thanks L.~R.~Werneck for useful discussions. N.J. and Z.B.E. gratefully acknowledge support from the NASA Astrophysics Theory Program under award No.~80NSSC22K1898 and the University of Idaho P3-R1 initiative. In addition, Z.B.E. acknowledges support from NSF awards PHY-2110352/2508377, PHY-2409654, OAC-2004311/2227105, OAC-2411068, OAC-2608811, and AST-2108072/2227080, as well as NASA award TCAN-80NSSC24K0100. This research made use of Idaho National Laboratory's High Performance Computing systems located at the Collaborative Computing Center and supported by the Office of Nuclear Energy of the U.S. Department of Energy and the Nuclear Science User Facilities under Contract No.~DE-AC07-05ID14517.
\end{acknowledgments}

\appendix
\section{Fisheye normalization and physical parameters}
\label{appendix:fisheye_scaling}

The forward fisheye map in Eq.~\eqref{eq:forward_fisheye_radial_map} uses transition centers $R_j$ and width parameters $s_j$ measured in computational radius. For grid construction, we instead prescribe the physical transition centers $R_{j,\rm phys}$, widths $w_{j,\rm phys}$, stretching factors $a_0,\ldots,a_n$, and physical boundary radius $L$ along each coordinate axis. We then determine the computational parameters $R_j$, $s_j$, and normalization $c$ that reproduce these physical inputs.

We choose the computational extent along each axis to be $[-L,L]$. Requiring $r(L)=L$ fixes the physical boundary locations and determines the normalization:
\begin{equation}
\begin{aligned}
c=L\Bigg[&a_nL+\sum_{j=1}^{n}\frac{(a_{j-1}-a_j)s_j}{2\tanh(R_j/s_j)} \\
&\times\ln\!\left(\frac{\cosh((L+R_j)/s_j)}{\cosh((L-R_j)/s_j)}\right)\Bigg]^{-1} \ .
\end{aligned}
\label{eq:fisheye_boundary_normalization}
\end{equation}
Thus $c$ is calculated directly from the boundary condition for any trial values of $R_j$ and $s_j$.

For each transition, the computational center $R_j$ must map to the prescribed physical center $R_{j,\rm phys}$. We define the physical width $w_{j,\rm phys}$ as the separation between the mapped endpoints of the computational interval $[R_j-s_j,R_j+s_j]$. These requirements give
\begin{equation}
\left\{
\begin{aligned}
r(R_j)&=R_{j,\rm phys} \ ,\\
r(R_j+s_j)-r(R_j-s_j)&=w_{j,\rm phys} \ ,
\end{aligned}
\right.
\quad j=1,\ldots,n \ .
\label{eq:fisheye_physical_conditions}
\end{equation}
The computational interval has width $2s_j$, while its mapped physical width is $w_{j,\rm phys}$. Because the map is nonlinear, the mapped endpoints are generally not symmetric about $R_{j,\rm phys}$.

Equation~\eqref{eq:fisheye_physical_conditions} supplies $n$ center conditions and $n$ width conditions for the $2n$ unknowns $R_j$ and $s_j$. These conditions are coupled because every transition contributes to the radial map and its normalization. We solve them simultaneously using a damped Newton method with a numerical Jacobian. For each trial set of $R_j$ and $s_j$, we first calculate $c$ from Eq.~\eqref{eq:fisheye_boundary_normalization}, then evaluate the differences between the mapped centers and widths and their prescribed values. We update $R_j$ and $s_j$ until these differences satisfy the convergence tolerance.

Away from well-separated transitions, $dr/d\hat r\simeq ca_j$, giving a physical grid spacing $\Delta x_{\mathrm{Cart},j}^{i}\simeq ca_j\Delta\hat x^i$ along coordinate axis $i$, where $j$ labels the layer. The stretching factors therefore determine the relative grid spacings, while $c$ sets their overall scale. For $N_{\hat x^i}$ grid points across $[-L,L]$ in computational direction $\hat x^i$, $\Delta\hat x^i=2L/N_{\hat x^i}$. With $a_0=1$, a target central spacing $\Delta x_{\mathrm{Cart},0}^{i}$ requires
\begin{equation}
N_{\hat x^i}\simeq\frac{2Lc}{\Delta x_{\mathrm{Cart},0}^{i}} \ .
\end{equation}

For the inverse map in Eq.~\eqref{eq:inverse_fisheye_radial_map}, the transition centers and widths are specified directly in physical coordinates, so no parameter conversion is required. We determine $c$ by enforcing $\hat r(L)=L$. Numerical inversion is used to obtain the physical positions of computational grid points.

For the inverse spheroidal map, we determine $c$ by requiring $\hat x=L$ at the physical point $(L,0,0)$. This fixes the computational equatorial extent to $[-L,L]$. Since the map stretches the polar and equatorial directions differently, the same computational extent in the polar direction would generally give a different physical boundary distance. We therefore evaluate the normalized inverse map at the physical point $(0,0,L)$ and use the resulting computational polar coordinate as the grid's polar half-extent. The physical boundaries then lie at $\pm L$ along each coordinate axis, while the computational polar and equatorial extents generally differ.

\bibliography{ref}
\end{document}